# Atomic Origin of Annealing Embrittlement in Metallic Glasses

Rui Su[1], Shan Zhang[2], Xuefeng Zhang[1,*], Yong Yang[3,4,*], Weihua Wang[5], Pengfei Guan[2,1,*]

[1]Institute of Advanced Magnetic Materials, College of Materials and Environmental Engineering, Hangzhou Dianzi University, Hangzhou 310018, P. R. China

[2]Beijing Computational Science Research Center, Beijing 100193, P. R. China

[3]Department of Mechanical Engineering, College of Engineering, City University of Hong Kong, Kowloon Tong, Kowloon, Hong Kong, China

[4]Department of Materials Science and Engineering, College of Engineering, City University of Hong Kong, Kowloon Tong, Kowloon, Hong Kong, China

[5]Institute of Physics, Chinese Academic of Science, Beijing 100193, P. R. China

**Corresponding authors.** XFZ (zhang@hdu.edu.cn), YY (yonyang@cityu.edu.hk) and PFG (pguan@csrc.ac.cn)

# Abstract

An atomistic understanding of annealing embrittlement is a longstanding issue for metallic glasses, which is still lacking due to the insurmountable gap between the thermal history of atomic models and laboratory-made samples. Here, based on a thermal-cycling annealing method that can vary the effective quenching rate over ten orders of magnitude, we perform an atomistic study of the ductile-brittle transition in a ternary model metallic glass, which can be keyed to the annealing embrittlement in bulk metallic glasses. We reveal that thermal annealing can effectively obliterate thermally-activable "defects", which are abundant in the hyper-quenched and ductile glass but gives rise to strain-created shear events in the well-annealed and brittle glass. While the activation of the strain-created events eventually causes single shear banding, other local structural disruptions can be "healed" by the same type of events upon stress reversal, thereby hindering shear band broadening or multiplication, and resulting in annealing embrittlement.

**150/150 Words**

# MAIN TEXT

## Introduction

Thermal annealing provides an effective way to regulate the macroscopic physical properties of metallic glasses (MGs), such as strength[1,2] and soft magnetic properties[3], so as to better meet the requirements of structural or functional applications. Annealing embrittlement, which is the phenomenon that fracture toughness of MGs deteriorates significantly with the annealing time[4–7], however is a stumbling block against comprehensive regulation of multi-performances. Several theoretical models have been proposed to understand this important phenomenon, such as the "free volume" model[8,9], which links the embrittlement to a reduction in "free volume" during thermal annealing. Due to the limited spatial and temporal resolution of current experimental techniques, the revelation of the atomic-level mechanism underpinning annealing embrittlement in MGs poses a severe challenge[10,11]. In the past two decades, molecular dynamic (MD) simulations[12–14] have been playing a crucial role in understanding the atomic structure of MGs and its response to external stimuli (e.g. mechanical force). However, conventional MD protocols can only afford a minimum quenching rate of $\sim 10^8$ *K/s*, which is at least 4 orders of magnitude faster than the maximum quenching rate of the copper casting technology commonly employed for experimental preparation of bulk metallic glass (BMGs)[15,16]. As a result, MD samples usually appear ductile because of this computational hyper-quenching, which hampers the understanding of the atomic-level mechanism underlying the annealing embrittlement in BMGs. We note that the recent works[17,18] demonstrated that the swap Monte Carlo (MC) method can effectively generate well-annealed polydisperse glasses in numerical simulations. While the efficiency of this method is generally sensitive to the dispersion of particles (e.g., ions, atoms), it is worthwhile to develop an effective method for MGs in order to overcome the seemingly insurmountable gap in the thermal history between a computationally simulated and laboratory-made MG.

## Results

**Thermodynamic characterization.** In order to effectively simulate thermal annealing on the MD platform, we develop a novel efficient annealing protocol in this work, termed as the Hybrid MD/MC Thermal Cycle (HTC) method hereafter, which combines conventional MD simulations with atomic swapping MC and sub-$T_g$ thermal cycling. We first validate this HTC method on a ternary model Zr-based MGs (See Methods and Supporting Information for more details). Note that the annealing efficiency of our HTC protocol is mainly contributed by the sub-$T_g$ thermal cycling, which accelerates local structural relaxations caused by the exchanges of solute atoms, either small (e.g. Cu) or large (e.g. Al) with solvent atoms (e.g. Zr). This behavior implies that structural relaxation in our simulated MG could be mainly due to local chemistry, in contrast to the size dependent structural relaxation in the previous swap MC simulations[17–19]. To characterize the degree of this computational "thermal annealing", we extrapolate the effective quenching rate $Q_e$ of the HTC-prepared samples by extrapolating the curve of the inherent structural energy per atom ($E_{IS}$) versus the computational quenching rate. Figure 1a shows the logarithmic dependence of $E_{IS}$ on the quenching rate $Q$ of samples prepared with MD in the range of $10^9 \leq Q \leq 10^{13}$ K/s, which is consistent with previous studies[20]. Evidently, by extrapolating the curve of $E_{IS}$ versus Q, the effective quenching rate $Q_e$ of our HTC-prepared MD samples can be obtained as low as $\sim 10^4$ K/s, which is comparable to the typical quenching rate in copper mold casting[15,16]. Here, we note that, as $Q_e$ reduces to $\sim 10^3$ K/s, partial crystallization can be observed in our HTC samples (Fig. S1a), which could be keyed to the critical cooling rate for vitrification (see calculated pair correlation functions $g(r)$ in Fig. S1b). In this work, the HTC samples with $Q_e \sim 10^4$ K/s correspond to the amorphous structure that can be obtained at the lowest cooling rate afforded by our inter-atomic potential[21], corresponding to real MGs obtained at its critical cooling rate.

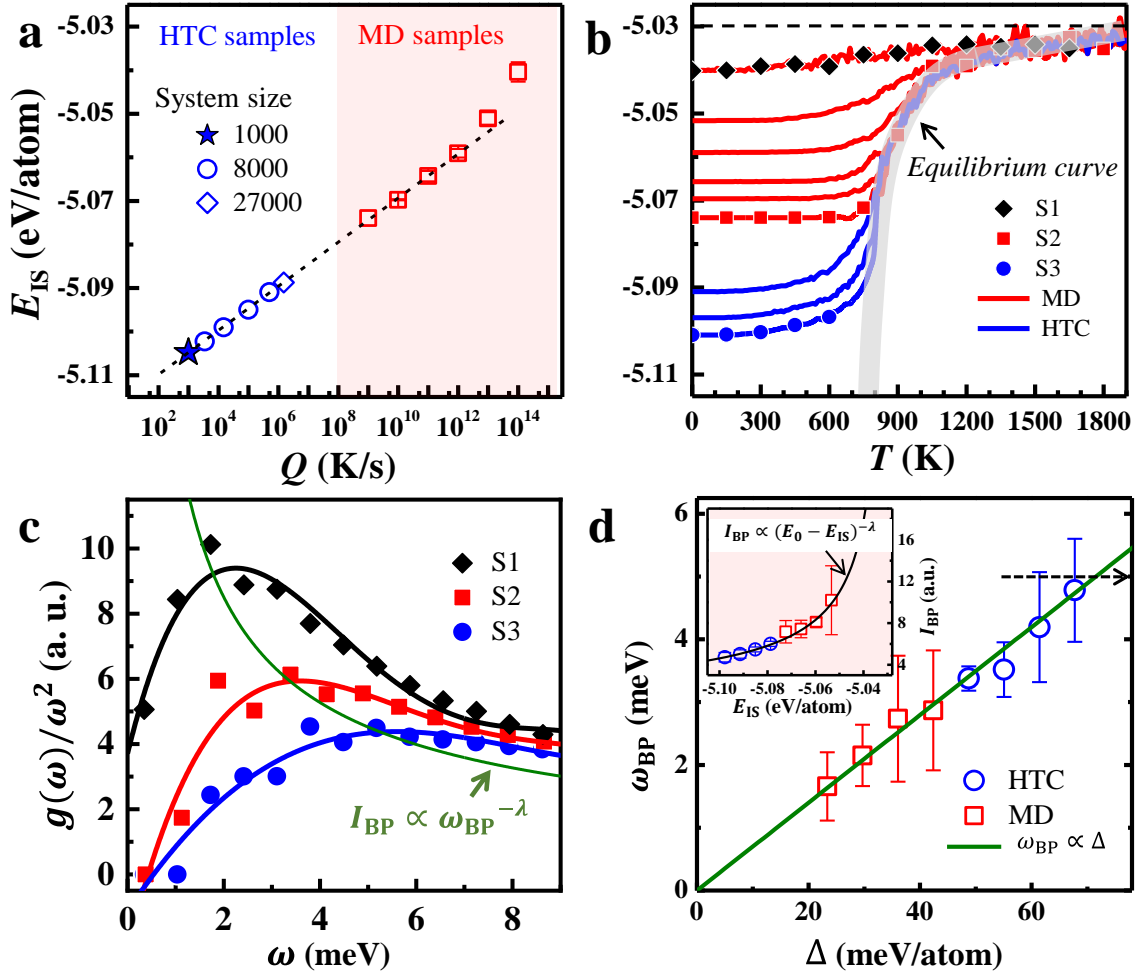

**Figure 1 | Thermodynamic characterization of the model metallic glass**. **a,** The potential energy per atom $E_{IS}$ of inherent structures, prepared by MD or HTC protocols, as a function of cooling rate $Q$. The effective cooling rate of the HTC samples (blue) are extrapolated by the linear fitting of the MD samples (read). The Error bars are calculated from ten independent runs. **b**, The evolution of $E_{IS}$ as the temperature increases for MD or HTC samples. The curve of the liquid-solid equilibrium is highlighted in the gray band and the upper limit of $E_{IS}$ is ~−5.03 eV/atom. Samples S1, S2 and S3 represent a set of samples with $Q \approx 10^{14}$, $10^9$ and $10^4$ K/s, respectively. **c,** The calculated $g(\omega)/\omega^2$ of samples S1, S2 and S3. The related Boson peaks can be observed and characterized by $\omega_{BP}$ and $I_{BP}$, which can be well fitted by $I_{BP} \propto \omega_{BP}^{-\lambda}$ with $\lambda = 0.69 \pm 0.09$. **d,** the $I_{BP}$ as a function of $E_{IS}$ (inset), which can be well fitted by $I_{BP} \propto (E_0 - E_{IS})^{-0.69 \pm 0.09}$ with $E_0 = -5.03$ eV/atom. The linear correlation of $\omega_{BP}$ with the annealing degree $\Delta = E_0 - E_{IS}$.

Figure 1b shows the evolution of $E_{IS}$ as a function of temperature for both MD and HTC simulations. Owing to the effective "annealing", the values of $E_{IS}$ obtained from the HTC simulations are well below those from the MD simulations in the low temperature regime, and both systems (e.g., HTC and MD) clearly deviate from the liquid-solid equilibrium. As a result of metastability, this behavior is consistent with the rate dependence of $E_{IS}$ shown in Fig. 1a. As temperature rises, the $E_{IS}$ values extracted from both HTC and MD simulations increase and finally merge onto the same master curve corresponding to the liquid-solid equilibrium that is obtained via high-energy state space sampling. Apart from the fact that the HTC simulations can probe a much lower energy state with smaller $E_{IS}$ on the potential energy landscape, we note that the glass transition temperatures of the HTC generated samples are less dependent on $E_{IS}$ (or cooling rate) than those of the conventional MD samples, which manifests in the trend of glass transition temperature versus cooling rate, as found in the previous MD studies[20]. Consequently, with the aid of both HTC and MD simulations, we successfully generated the atomic model for the same MG with the effective quenching rate spanning over ten orders of magnitude ($10^{14}$ to $10^4$ K/s). This extremely large temporal window is unprecedented, offering us an opportunity to investigate the atomic mechanisms of thermal annealing induced embrittlement in BMGs that cannot be done otherwise.

To further validate our simulations through the comparison with experiments, we examine the Boson peaks of the Zr-based MG using our MD and HTC simulations. Here, we note that through the prior expriments[22,23], it is known that the Boson peaks are correlated with the energy stability and macroscopic physical properties of MGs. By calculating the state of vibrational density $g(\omega)$, the Boson peak, arising from the low-frequency vibrational modes that exceed the Debye's prediction, can be characterized as the peak on the curve of $g(\omega)/\omega^2$ versus $\omega$. Figure 1c shows our calculated $g(\omega)/\omega^2$ against $\omega$ for three representative samples (S1, S2, and S3) with $Q_e \sim 10^{14}$, $10^9$, and $10^4$ K/s respectively. As $Q$ or $E_{IS}$ is decreased by ten orders of magnitude, there is a dramatic shift of the frequency ($\omega_{BP}$) of the Boson peak to a higher energy (frequency), alongside a significant reduction in the peak intensity ($I_{BP}$). According to the Euclidean random matrix theory[24] and the previous numerical results[25], $\omega_{BP}$ and $I_{BP}$ respectively follow the power-law relations: $\omega_{BP} \propto \Delta$ and $I_{BP} \propto \Delta^{-\beta}$ ($\beta < 1$), where $\Delta$ is the parameter representing the energy stability of glass, which leads to $I_{BP} \propto \omega_{BP}^{-\beta}$. We plot the data of $E_{IS}$ versus $I_{BP}$ for all our samples (including MD and HTC samples) in the inset of Fig. 1d, which can be well fitted to $I_{BP} \propto (E_0 - E_{IS})^{-0.7}$ with $E_0 = -5.03$ eV/atom. Since $E_0$ corresponds to the instability limit of an amorphous elastic phase ($I_{BP} \to \infty$), and $\Delta \equiv -5.03 - E_{IS}$ can be defined as the annealing degree of each sample with respect to the upper limit of $E_{IS}$ (~$-5.03$ eV/atom, Fig. 1b), a perfect linear relation between $\Delta$ and $\omega_{BP}$ can be established in Fig. 1d for all our samples investigated. Furthermore, the $\omega_{BP}$ of sample S3 with $\Delta \sim 70$ meV/atom is estimated to be ~5 meV, which is in excellent agreement with the inelastic neuron scattering experiments[26] on typical BMGs. This evidence delivers a strong message that our well-annealed HTC samples indeed approach the energy state of BMGs.

**Annealing embrittlement.** Figure S2a-d show the calculated bulk modulus $B$, shear modulus $G$, mass density $\rho$, and Poisson's ratio $\nu$ of our numerical samples, which vary linearly with the annealing degree $\Delta$, being consistent with the prior experimental findings[27]. The significant decrease of Poisson's ratio (ν) with $\Delta$ implies that the toughness of our samples may deteriorate significantly with annealing time[27,28]. To study the yielding behavior of these samples, athermal quasi-static shearing (AQS, see Methods) loadings were performed, and the ensemble-averaged (based on 100 replicas with the same $\Delta$) stress-strain curves for representative $\Delta$ are shown in Fig. 2a. The shape of the curve presents a qualitative change from a smooth crossover (ductile-like behavior) for the poorly annealed S1 samples ($\Delta \approx 20$ meV/atom) and S2 samples ($\Delta \approx 45$ meV/atom) to a sharp discontinuity (brittle-like behavior) for the well-annealed S3 samples ($\Delta \approx 70$ meV/atom, see in Fig. S3). For a given $\Delta$, the extracted maximum stress drops $\Delta\sigma_{max}$ and the corresponding strains $\gamma_{max}$ at yielding of 100 replicas are collected. As $\Delta$ increases, the distribution of $\gamma_{max}$ becomes narrowed, accompanied by a monotonic increase in $\langle\Delta\sigma_{max}\rangle$ (Fig. 2b). The funnel-like shape of the distribution indicates that the yielding strain converges to a single value with the increase of $\Delta$. This behavior is intriguing. On the one hand, it is known that thermal annealing can help sample different atomic configurations; on the other hand, thermal annealing results in similar yield points and a sharp discontinuity on the stress-strain curve once yielding occurs, as seen in Fig. 1a. To characterize the brittle-ductile transition, we studied the distribution of stress drop $\Delta\sigma_{max}$. The distribution width of $\Delta\sigma_{max}$ is peaked at an intermediate $\Delta$ while the average value $\langle\Delta\sigma_{max}\rangle$ monotonously increases with $\Delta$ (Fig. S4a). As a result, the calculated variance of $\Delta\sigma_{max}$: $\chi_{max} = \langle\Delta\sigma_{max}^2\rangle - \langle\Delta\sigma_{max}\rangle^2$ exhibit a sharp peak (Fig. S4b) at a critical annealing degree: $\Delta_c \approx 60$ meV/atom ($Q_c \approx 10^6$ K/s). Notably, a similar behavior was observed in other glass formers when they undergo a brittle-ductile transition[18]. Therefore, $\Delta_c \approx 60$ meV/atom can be viewed as a critical annealing degree that separate the intrinsic ductile and brittle behaviors in our HTC simulations, which is far beyond the capability of conventional MD simulations.

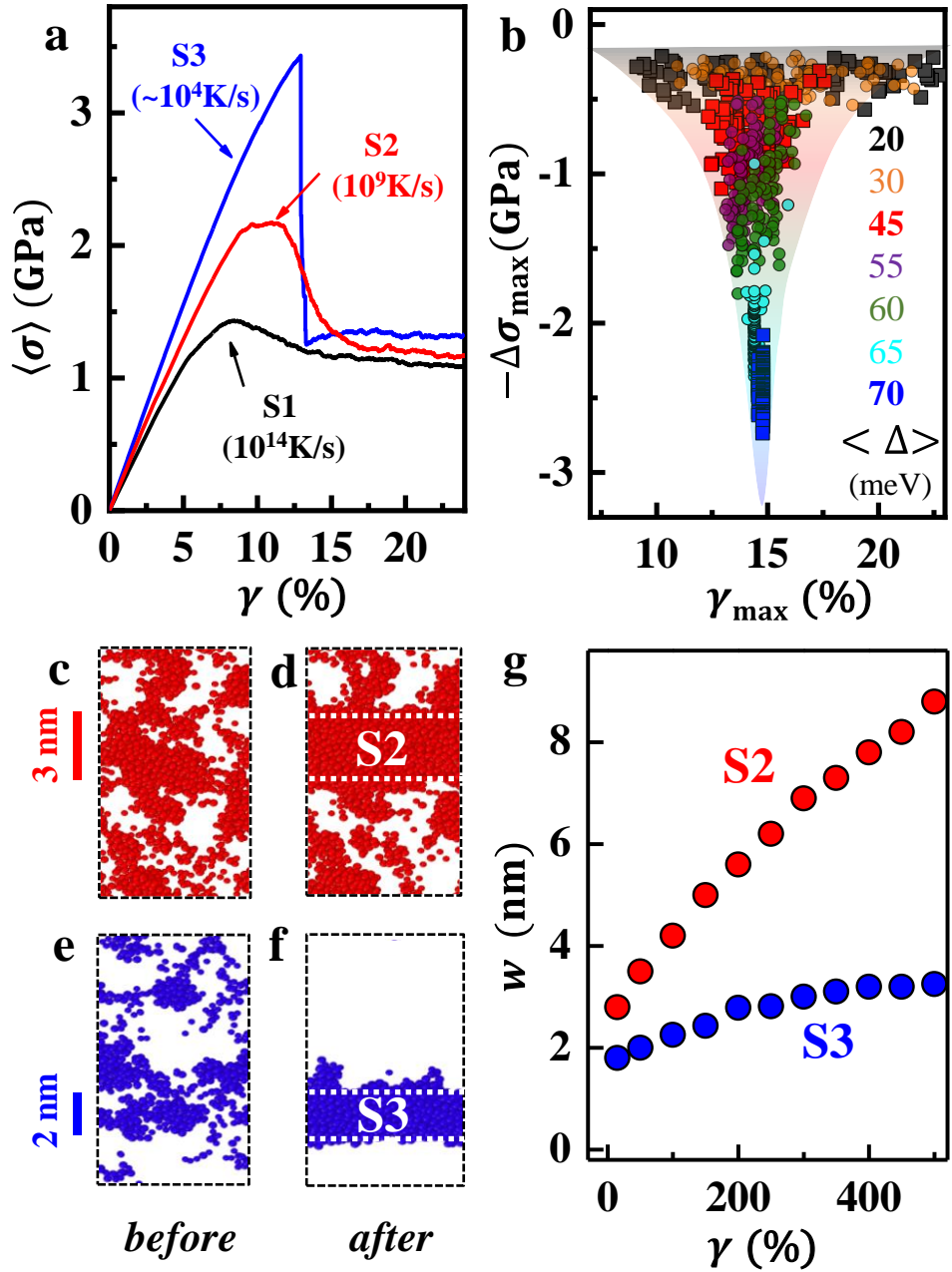

**Figure 2 | Annealing embrittlement revealed by the atomistic simulations**. **a,** Calculated stress-strain curves for samples **S1**, **S2**, and **S3** exhibit distinct yielding behaviors. **b,** The maximum stress drops $\Delta\sigma_{max}$ and the corresponding strains $\gamma_{max}$ at yielding of 100 replicas for each given $\langle\Delta\rangle$. **c-f,** Spatial distributions of local atomic rearrangements (atoms with

$D^2_{\min}$>10) just before (left panel) and after (right panel) yielding for samples **S2** and **S3**. **g,** The width (*w*) of shear band as a function of the external shear strain.

Atomic non-affine displacement ($D^2_{\min}$)[29] just before and after yielding, with respect to the initial configuration are calculated to represent distinctive atomic-level structural rearrangements in the samples with different Δ. For sample S1 ($\ll \Delta_c$), there is a uniform spatial distribution of shear transforming (ST) events with large $D^2_{\min}$ (>10, atoms in black) before yielding (Fig. S4c) and their post-yielding connections, which manifests as the formation of an isotropic percolation (Fig. S4d), indicate a homogenous plastic deformation. As a result, we do not observe the formation of any visible sample-spanning "shear bands". By comparison, sample S2 exhibits localized activations of ST events (Fig. 2c), which evolve into a shear band by coalescence upon yielding (Fig. 2d), being consistent with the previous MD works[30,31]. Furthermore, the rearranged sites outside the shear band region survive from the stress-releasing induced by the shear band formation. In sharp contrast, sample S3 forms an extremely thin (with a thickness of ~2.0 nm) shear band following the activations of local rearrangements, but the severely rearranged sites outside the shear band region seemingly become "healed" entirely (Fig. 2e-f), indicative of a local structural recovery from rejuvenated states with the shear band formation. It is noteworthy that such a post-shear structural healing has never been reported in the literature even through shear banding has been studied extensively in the previous experimental and theoretical works[32,33]. The dynamics of shear banding, including its nucleation and propagation that govern the failure behaviors of MGs, is sensitive to the pre-rejuvenated zones around the front of a growing shear band. Following this line of reasoning, one can envision that the post-shear structural healing in sample S3 restricts shear band broadening, as seen in Fig. 3g. Furthermore, it is noteworthy that the width (*w*) of the thin shear band in sample 3 is insensitive to the external shear strain (see Fig. S5a). In stark contrast, the width of the shear band in sample S2 increases rapidly[33] with the external shear strain (see Fig. S5b).

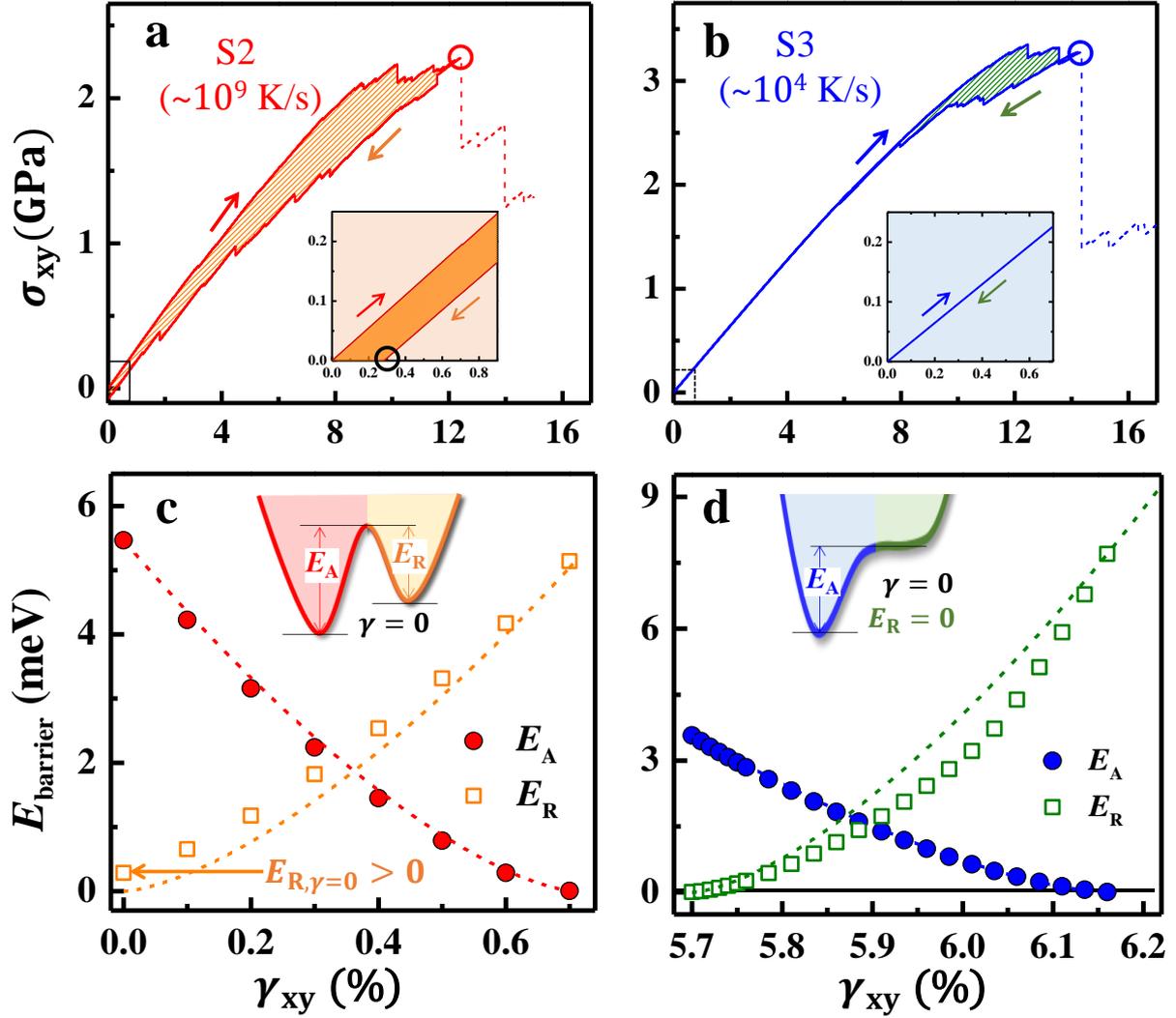

**Figure 3 | Thermally-activable versus strain-created ST events.** The calculated hysteretic cyclic stress-strain curves subjected to the external strain $\gamma_{xy}$ until yielding occurs in sample S2 (**a**) and S3 (**b**). The inset shows the enlarged view of the region close to $\gamma_{xy} = 0$. The calculated activation ($E_A$) and relaxation $E_R$ barriers against external strain $\gamma_{xy}$ for a typical event derived from the mechanical hysteresis in sample **S2 (c)** and **S3 (d)**. Note that the dashed curves are the prediction of the catastrophe theory: $E(\gamma) = A(\gamma_c - \gamma)^{3/2}$, where A is a constant pre-factor and $\gamma_c$ is the critical strain triggering a catastrophe event.

**Shear transforming events.** To understand the local structural recovery with the shear band formation, we unloaded the samples from their yielding point to zero stress. Figures 3a and 3b present the loading-unloading stress-strain curves for samples S2 and S3, respectively.

Hysteresis loops can be observed in both samples due to the local rearrangements induced internal friction inherent to amorphous solids[30,33]. The close loop in sample S3, in contrast to the open loop in sample S2, hints that all the energy dissipating rearrangements are reversible in sample S3. This supports our observation that the well annealed amorphous structure is shear resilient and even appear to be "healable" at the local rearranged sites upon stress reversal (Fig. 2h-i). Figures 3c and 3d respectively show the calculated activation energy barrier $E_A$ and the relaxation barrier $E_R$ as a function of the external shear strain $\gamma$ for a typical event in two samples. It is evident that $E_A$ continuously reduces to zero as $\gamma$ approaches the critical strain $\gamma_c$ for both events, which is in good agreement with the catastrophe theory[34]. However, the ST event in sample S2 corresponds to $E_R > 0$ at $\gamma = 0$, while that in sample S3 to $E_R = 0$ at $\gamma > 0$, which implies that there are two types of ST events: one can be viewed as a strain-driven transition between two adjacent sub-basins separated by a saddle on the unperturbed potential energy surface (PES) ($\gamma = 0$) (Fig. S6a), which is thermally-activable; and the other can be interpreted as the transition between adjacent sub-basins on the strain-coded PES, which is strain-created and undetectable on the unperturbed PES (Fig. S6b). The results in Fig. 3a and 3b indicate that all ST events in sample S3 are strain-created, while part of the ST events in sample S2 are strain-created and the rest are thermally-activable. Therefore, the strain-created events are responsible for the brittle behavior of the well-annealed samples: on the one hand, they are the embryos to nucleate an extremely thin shear band; on the other, they restrict shear band broadening and/or multiplication by elastic shielding.

**Low-frequency modes and soft spots.** The atomic structures of the rearranged sites are characterized by the local atomic packing analysis. As shown in Fig. S7, atom rearrangements are more likely to occur at the sites with a low five-fold symmetry[35] in the hyper-quenched and ductile sample S2, but appear sporadic in the well-annealed and brittle sample S3, which, however, lack a clear geometric identity. Alternatively, we utilized atomic softness[36] $S_i$, which was defined based on quasi-localized low-frequency eigenmodes (see Methods), to quantify "soft spots" or the fertile sites for ST events. Here, we note that atomic softness was considered in the literature as the defining feature of "flow defects" or "flow units" [37,38] to initiate ST events[39–41] in MGs under mechanical stimuli. Figures 4a and 4b present the participation ratio[36] $P_i$ (see Method) of thermal vibration modes in samples S2 and S3. A similar behavior was observed in other glass formers[36] that the quasi-localized modes with small $P_i$ values are reduced in number and significantly shifted to higher frequencies in well-annealed and brittle sample S3. Differently, the brittle sample exhibits the lowest $P_i \approx 0.01$, which indicates that the most localized low-frequency mode involves ~100 atoms. However, the most localized mode in the ductile sample (S2) involves only ~10 atoms.

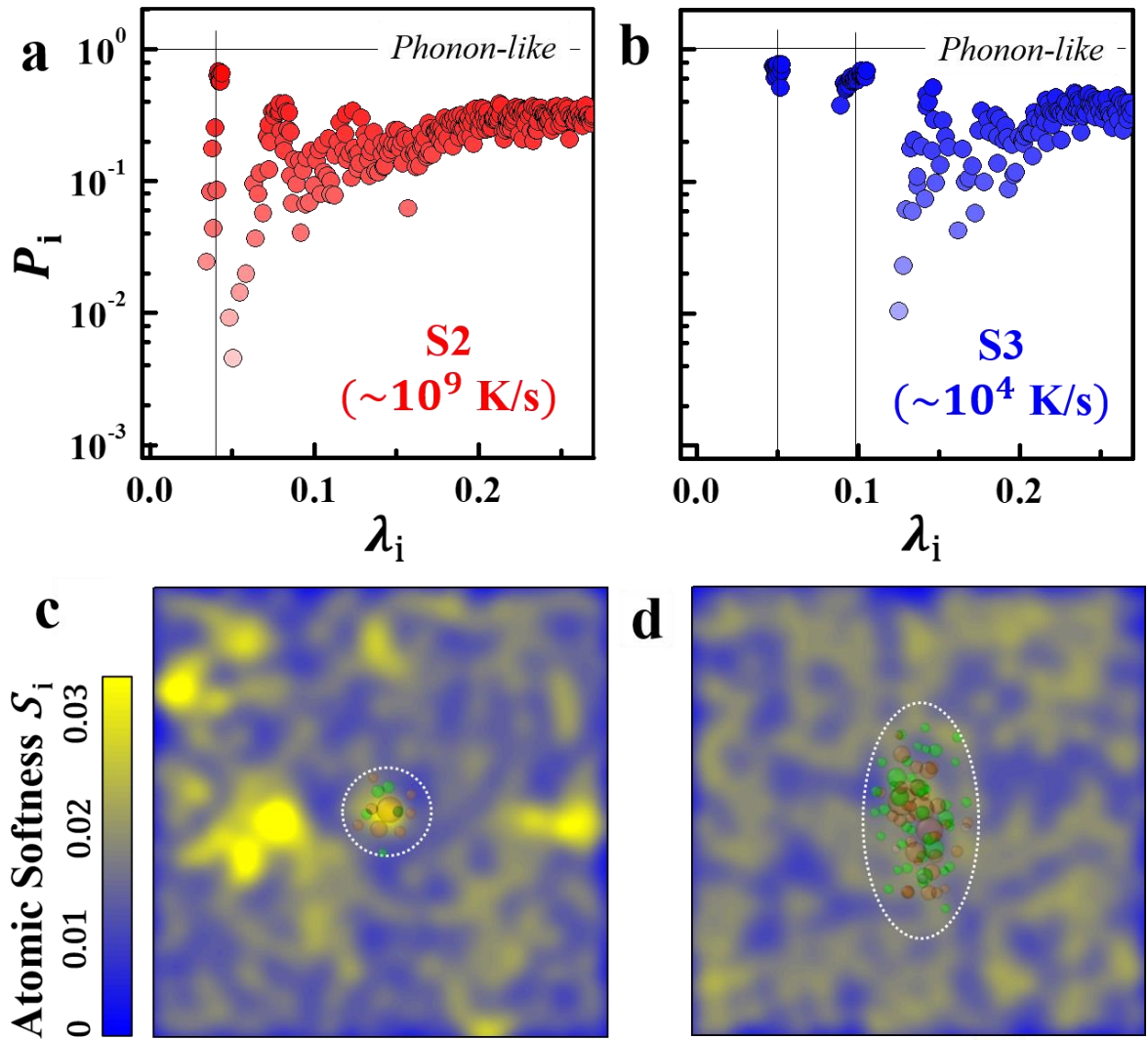

**Figure 4 | Characterization of thermal vibrations and soft spots**. **a-b**, Participation ratio $P_i$ of eigen modes in sample **S2** and **S3**, respectively. Spatial distributions of atomic softness $S_i = 1/N_m \sum_{i \leq N_m} |\vec{e_i}|$, which is calculated based on the vibrational modes of the inherent structure of sample **S2 (c)** and **S3 (d)**. Soft spots are colored in yellow, and atoms involved in the first event are encircled.

As a result, one can see in Fig. 4c that the distribution of $S_i$ is rather inhomogeneous in sample S2, in which the local regions with large $S_i$ may be interpreted as the "soft spots". In such a case, we observed that the first ST event in sample S2 did occur at one of the "soft spots", which agreed well with the previous MD simulations[39,40]. However, the $S_i$ distribution in sample S3 looks relatively homogenous (Fig. 4d) and a clear distinction between "soft spots" and their surroundings are smeared, hence defying any meaningful identification of the first ST event on the basis of "soft spots". Moreover, a typical ST event in sample S3 corresponds to a larger activation volume containing ~100 atoms with an ellipsoid-like morphology (Fig. 4d), in contrast to a spherical-like activation volume containing ~10 atoms (Fig. 4c) for the event in sample S2. This finding indicates that the activation volume of the ST events increases with Δ and this behavior may reconcile the long-standing discrepancy between the shear transformation zone (STZ) size (~10 atoms) extracted through MD simulations[12,13,42] and the experimental estimations[43–45] (~100 atoms) since the thermal history of a sample can alter the STZ size.

**Potential energy surface.** From the viewpoint of PES, the thermally-activable defects can be identified a priori by the sub-basins in a meta-basin and it is the heights of the sub-basins that determine the activation energies ($E_A$, see Methods) of the thermally activated transition events. As shown in Fig. 5a, the $E_A$ spectrum "blueshifts" and the number of thermally-activable events decreases greatly with Δ. Note that the magnitude of $E_A$ for the individual event is less than a typical $β$ relaxation energy (~0.6 eV)[43,44], which implies that the low-temperature structural relaxation induced by thermal activation is ultra-weak in the brittle samples with $Δ ≈ 70 \text{meV/atom}$. Moreover, in our ductile samples (e.g. S2), plenty of thermally-activable defects can be activated at a critical strains[12] ($γ_c$) much less than the overall yielding strain ($γ_0$, Fig. 5b). Hence, ductile type yielding (or shear banding) is due to the accumulation of local thermally-activable events. However, there are no such thermally-activable defects that can be triggered in our brittle sample (e.g. S3, Fig. 5c) before overall yielding. A ductile system should be located at a meta-basin populated with sub-basins, and yields via the accumulation of the activation of local thermally-activable defects. However, thermal annealing can drive the system into a deeper and smoother meta-basin as illustrated in Fig. 4d, and strain-created events prevail in the well annealed system.

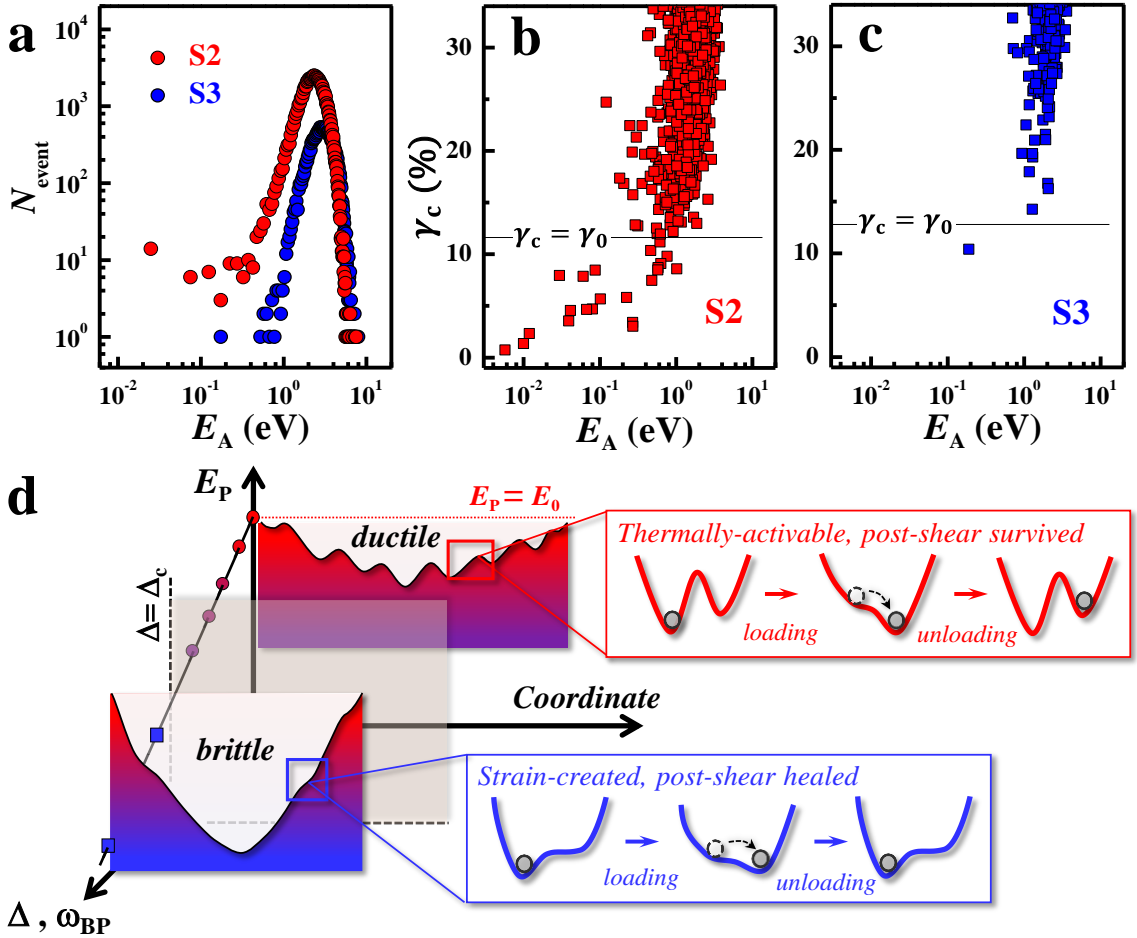

**Figure 5 | The spectra of activation barrier $E_A$ and the illustrated potential energy surface (PES). a,** Thermal activation barrier ($E_A$) spectra of sample **S**2 and **S**3. **b-c,** The predicted critical strains $\gamma_c$ of the insular thermally-activable events between neighboring sub-basins with various $E_A$ in samples S2 and S3, respectively. The yielding strains $\gamma_0$ of each sample are illustrated as thin solid lines. **d,** Schematic diagram of the local PES evolution that determines thermal-annealing embrittlement.

## Discussion

In summary, based on the above MD and HTC simulations, we extracted the distributions of maximum stress drop $\Delta\sigma_{max}$ and corresponding strain $\gamma_{max}$ from the simulated strain-stress curves, which signal the ductile-brittle transition induced by thermal annealing that cannot be observed in conventional MD simulations. Atomic-scale evidence reveals that shear banding can induce post-shear local structural recovery or "self-healing" at the strain-created rearranged sites outside the shear band region in the brittle samples, a phenomenon accompanied by the formation of an extremely thin shear band that does not widen with the applied shear strain. Here, we emphasize that, unlike thermally-activable ST events that can be linked to sub-basins, these "healed" ST sites are strain created, which signals a local alteration of the PES topography. Therefore, as thermal annealing drives an amorphous system into a smooth, quasi-elastic meta-basin, only sporadic atomic rearrangements of a strain-created nature can be triggered prior to overall yielding. The lack of pervasive thermally-activable defects then leads to annealing embrittlement.

Furthermore, the developed HTC protocol allows the preparation of MG models corresponding to the thermal stabilities relevant to experiments, which bridges the gap between atomistic simulations and experimental studies. Systematic characterizations of structural and physical properties evolution over an unprecedented range of energy stabilities of MGs provides atomistic insights into the fundamental physics of BMGs. For example, as the thermal stability of model samples approaches those of real MGs, the previously assumed linear relationship between potential energy and the fraction of locally favorable structures, such as icosahedron in Cu-Zr-based MGs, no longer holds. Thus, it brings us new arguments on whether LFS provides the appropriate characterization of MGs stabilities, as mentioned in previous unsuccessful efforts to connect LFSs with dynamical arresting in MGs[42]. Instead, the validated uniform correlation between glass stability and Boson peak properties suggests that excess low-frequency vibrational modes may be the key to solving the problems of glass formation and mechanical stability of MGs.

## Methods

**Sample preparation.** A series of ternary ZrCuAl metallic glass samples were prepared by the classical molecular dynamic (MD) or the hybrid MD/MC thermal-cycling (HTC) method. All simulations were performed with the LAMMPS[46] package for $Cu_{46}Zr_{46}Al_8$ (at. %) metallic glasses with realistic embedded-atomic-method potentials[21]. The MD time step was set to 2 fs, and a constant pressure and temperature (NPT) ensemble was employed in which the temperature and pressure were controlled by the Nosé-Hoover[47,48] and Parrinello-Rahman methods[49] respectively. For the HTC simulations, Monte Carlo (MC) swapping[50] between atoms with different chemical elements was introduced for each MD step during the quenching process. For each atom swapping attempt, the atom velocities were scaled to conserve the total kinetic energy. The atom swapping was accepted by the Metropolis criterion: $P = \min\{1, \exp[-\beta \Delta U]\}$ where $\beta = 1/k_\text{B}T$ is the Boltzmann factor and the temperature $T$ was set to be equal to the MD temperature. For each MD step, 10 swap MC steps were performed.

Model systems containing 27000, 8000, and 1000 atoms in a cubic box with periodic boundary conditions were prepared to eliminate the size effect. We first kept the system at 2500 K (far above the melting point) for 10 ns. The MD samples were prepared by quenching the melt at selected rates ($10^{14}$, $10^{13}$, $10^{12}$, $10^{10}$ or $10^9$ K/s) from 2500K to 1K based on the classical MD simulations. For HTC samples, the melt was first quenched from 2500K to 700 K by the classical MD method and then subjected to thermal cycling between 700K to 1K based on the HTC simulations. The samples at 1K after different cycles were collected for further analysis.

**Thermal vibrational behaviors and atomic softness.** The vibrational density of states was computed by diagonalizing the dynamical matrix of the inherent structure. The inherent structure was obtained by instantaneously quenching the equilibrium configuration to the potential energy minimum and the dynamical matrix was calculated by disturbing the atom positions by small displacements. Matrix diagonalization was performed by a full diagonalization through the Intel Math Kernel Library (Intel MKL) library (https://software.intel.com/en-us/mkl/).

The spatial localization of the $i$-th eigenmode $e_i$ was defined as: $P_i = (\sum |e_i|^2)^2 / N \sum |e_i|^4$, where $e_i$ is the $i$-th eigenvector of Hessian matrix. The atomic softness is defined as: $S = 1/N_m \sum_{i<N_m} |e_i|$ and $N_m$ is the number of lowest modes and $e_i$ is the eigenmode of the Hessian matrix. The soft spots are defined as the local maximum of the density plot of per-atom softness[39,40]. In this work, we use $N_m = 50$ while changing $N_m$ from 30 to 100 does not affect our results.

**Mechanical testing and characterization of yielding behaviors.** Athermal quasi-static shear (AQS) simulations were performed by incrementally deforming the simulation cell of MG samples along the *xy* direction with a step strain $\Delta \gamma = 1 \times 10^{-5}$. Energy minimizations were performed with a force tolerance of $10^{-4}$ eV/Å. All samples display linear-elastic behavior at low strain, followed by a transition to the flow state at their yield stress. To measure the shear band broadening, the related sample is deformed in excess of 500% strain.

To identify the distinct yielding behaviors of samples with representative Δ, the overlapping function $Q_{12}$ between deformed and initial configurations, which is defined as:

$$Q_{12} \equiv \frac{1}{N}\sum_{i}^{N} \theta(a - |r_i^1 - r_i^2|) \quad (1)$$

is calculated at different external shear strain. Here, $N$ is the number of atoms, $r_i^1$ and $r_i^2$ are positions of atom $i$ in deformed and initial configurations, respectively. The $\theta(x)$ is the Heaviside step function, $a = 0.4$ Å is the distance cutoff for overlap checking and $\langle \cdot \rangle$ stands for ensemble average. Ensemble averaging is performed in a 100-samples ensemble that is construct by random perturbations. The dynamical susceptibility $\chi_{12}$ is defined as: $\chi_{12} \equiv N(\langle Q_{12}^2 \rangle - \langle Q_{12} \rangle^2)$.

The stress drops along the stress-strain curves are detected by the local tangent $d\sigma/d\gamma \approx (\sigma_i - \sigma_{i-1})/\Delta\gamma$ at each strain $\gamma_i$. The stress drop sizes are calculated as $\Delta\sigma = \sigma_a - \sigma_b$, where $a$ marks the strain point that $d\sigma/d\gamma$ changes from positive to negative and $b$ marks the point that $d\sigma/d\gamma$ become positive again after $a$. For a given Δ, the maximum stress drops $\Delta\sigma_{\max}$ and the related $\gamma_{\max}$ of 100 samples are collected.

**Rearranged atoms and Local atomic packing analysis.** The atomic rearrangements of MGs samples under external strain were monitored by the local minimum nonaffine displacement $(D_{\min}^2)^{29}$ of each atom. We compute the $D_{\min}^2$ of deformed states with the initial, undeformed configuration is selected as the reference state. A cutoff $D_{\min}^2$ of 10 Å² (Fig. S8) is used to select the rearranged atoms. The shear band width ($w$) is simply defined by measuring the thickness of the percolated region in the post-yielding state.

Voronoi tessellation divides space into close-packed polyhedral around atoms by constructing bisecting planes along the lines joining the central atom and all its neighbors[51]. The Voronoi index $\langle n_3, n_4, n_5, n_6 \rangle$ is used to characterize the geometry feature of atomic clusters, where $n_k$ ($k = 3,4,5,6$) denotes the number of $k$-edged faces of a Voronoi polyhedron. Based on the Voronoi analysis, we define the local five-fold symmetry parameter of atom $i$ as $f_{5,i} = n_{5,i}/\sum_{k=3,4,5,6} n_{k,i}$, where $n_{k,i}$ specifies the number of $k$-edged polygons in the Voronoi polyhedron centered at atom $i$. The $P_{\text{All}}(f_5)$ and $P_{\text{RA}}(f_5)$ are the fractions of different $f_5$ for all and rearranged atoms, respectively. The enhanced factor $\lambda$ of each type $f_5$ is defined as $\lambda(f_5) = P_{\text{RA}}(f_5)/P_{\text{All}}(f_5)$. If $\lambda(f_5) > 1$, the local structural rearrangements are more likely to occur at atoms with the special local five-fold symmetry.

**Minimum energy path, activation energy barrier and trigger strain of ST events.** To reveal the evolution of the activation energy barrier $E_A$ during the loading and unloading processes, the external strain-dependent full minimum energy path of the identified shear transforming event was calculated by the climbing image nudged elastic band (CI-NEB)[52] method. The CI-NEB calculations were performed with a force tolerance of $1 \times 10^{-4}$ eV/Å and 40 intermediate images.

The activation-relaxation technique[53] (ART) was employed for searching for the first-order saddle points of a given MG sample. The initial "activation" process of the system was performed by random displace atoms in a radius of 3.8 Å centering around a chosen atom. The subsequent "relaxation" process was accomplished by first converging to the nearby saddle point with a force tolerance of $5 \times 10^{-3}$ eV/Å and then converging to the nearby energy minimum with a force tolerance of $1 \times 10^{-4}$ eV/Å. For each saddle point, we constructed a "feature vector" containing energy barrier, lowest curvature, atom displacement magnitude and the stress components at saddle point which was subsequently used to calculate the Pearson's correlation P of a pair of saddle points and the redundant saddle points were removed by a correlation cutoff of P > 0.9998.

The triggering shear strain of ST event is calculated by the equation: $\gamma_c = -3Q_0/2V\Delta\tau_{0,\alpha\beta}$ as derived from analysis by Xu *et. al.*[12]. Here $Q_0$, $V$ and $\Delta\tau_{0,\alpha\beta}$ are energy barrier, system volume and stress component change at saddle point for each activation events predicted by ART calculations.

**Supplementary contents**

Extended data: structure information, Macroscopic properties, Yielding and shear banding broadening behaviors of model metallic glasses.

**Competing interests:** The authors declare that they have no competing interests.

**Data and materials availability:** All data needed to evaluate the conclusions in the paper are present in the paper and/or the Supplementary Materials. Additional data related to this paper may be requested from the authors.

**Acknowledgement**

This work is supported by the NSF of China (Grant Nos. 5211101002 and U1930402) and the National Key R&D Program of China (Grant No. 2017YFA0303400). R.S. acknowledges the Young Scientists Fund of the National Natural Science Foundation of China (No. 51801046). Y.Y. acknowledges the financial support from the University Grant Council through the General Research Fund (GRF) with grant numbers N_CityU109/21, CityU11213118 and CityU11209317. P.F.G. and S.Z. acknowledge the computational support from the Beijing Computational Science Research Center (CSRC).

**Author contributions**

P.F.G. supervised the work. R.S., S.Z., and P.F.G. performed modeling and simulations. R.S., X.F.Z, Y.Y., and P.F.G. wrote the manuscript. All the authors contributed to the analysis and interpretation of the data and to the development and editing of the manuscript.

# Figures

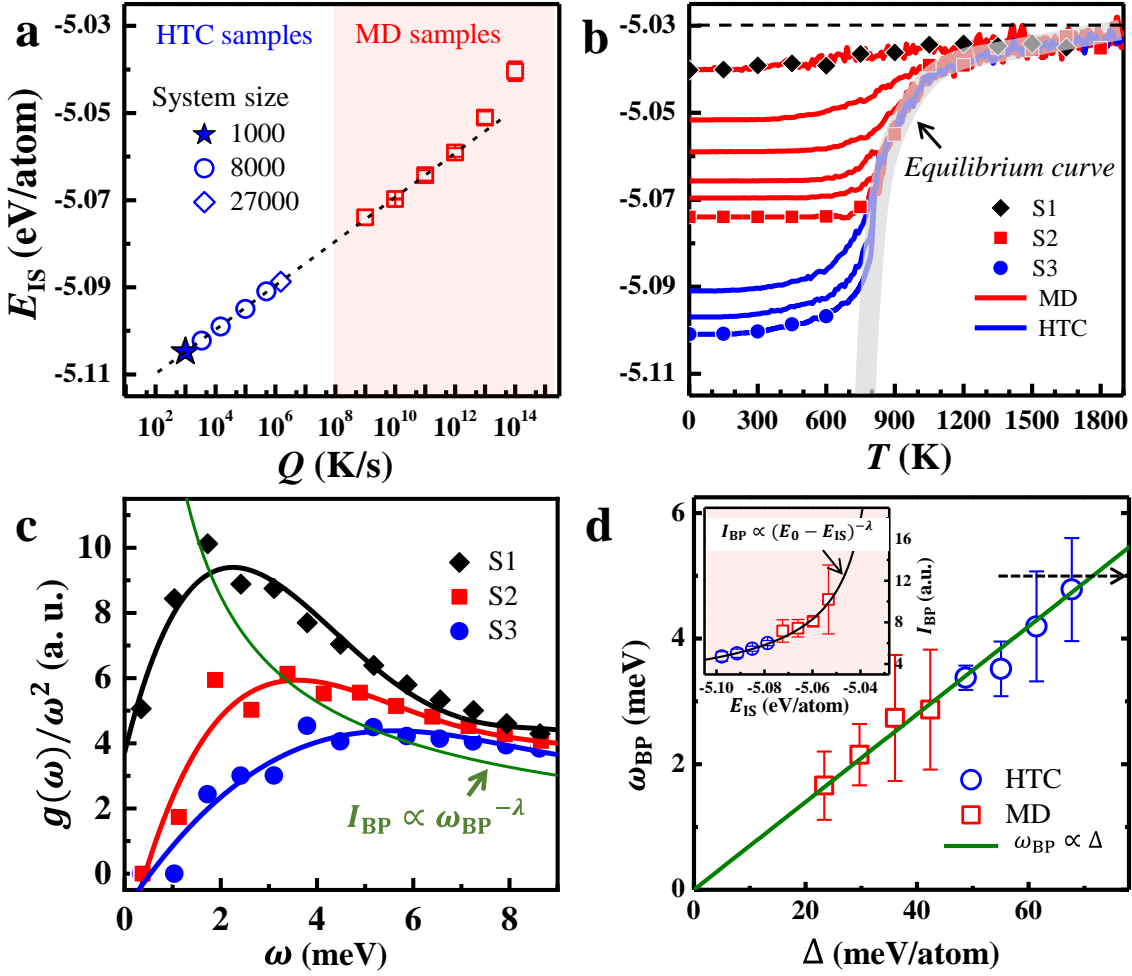

**Figure 1 | Thermodynamic characterization of the model metallic glass**. **a,** The potential energy per atom $E_{IS}$ of inherent structures, prepared by MD or HTC protocols, as a function of cooling rate $Q$. The effective cooling rate of the HTC samples (blue) are extrapolated by the linear fitting of the MD samples (read). The Error bars are calculated from ten independent runs. **b**, The evolution of $E_{IS}$ as the temperature increases for MD or HTC samples. The curve of the liquid-solid equilibrium is highlighted in the gray band and the upper limit of $E_{IS}$ is $\sim -5.03$ eV/atom. Samples S1, S2 and S3 represent a set of samples with $Q \approx 10^{14}$, $10^9$ and $10^4$ K/s, respectively. **c,** The calculated $g(\omega)/\omega^2$ of samples S1, S2 and S3. The related Boson peaks can be observed and characterized by $\omega_{BP}$ and $I_{BP}$, which can be well fitted by $I_{BP} \propto \omega_{BP}^{-\lambda}$ with $\lambda = 0.69 \pm 0.09$. **d,** the $I_{BP}$ as a function of $E_{IS}$ (inset), which can be well fitted by $I_{BP} \propto (E_0 - E_{IS})^{-0.69 \pm 0.09}$ with $E_0 = -5.03$ eV/atom. The linear correlation of $\omega_{BP}$ with the annealing degree $\Delta = E_0 - E_{IS}$.

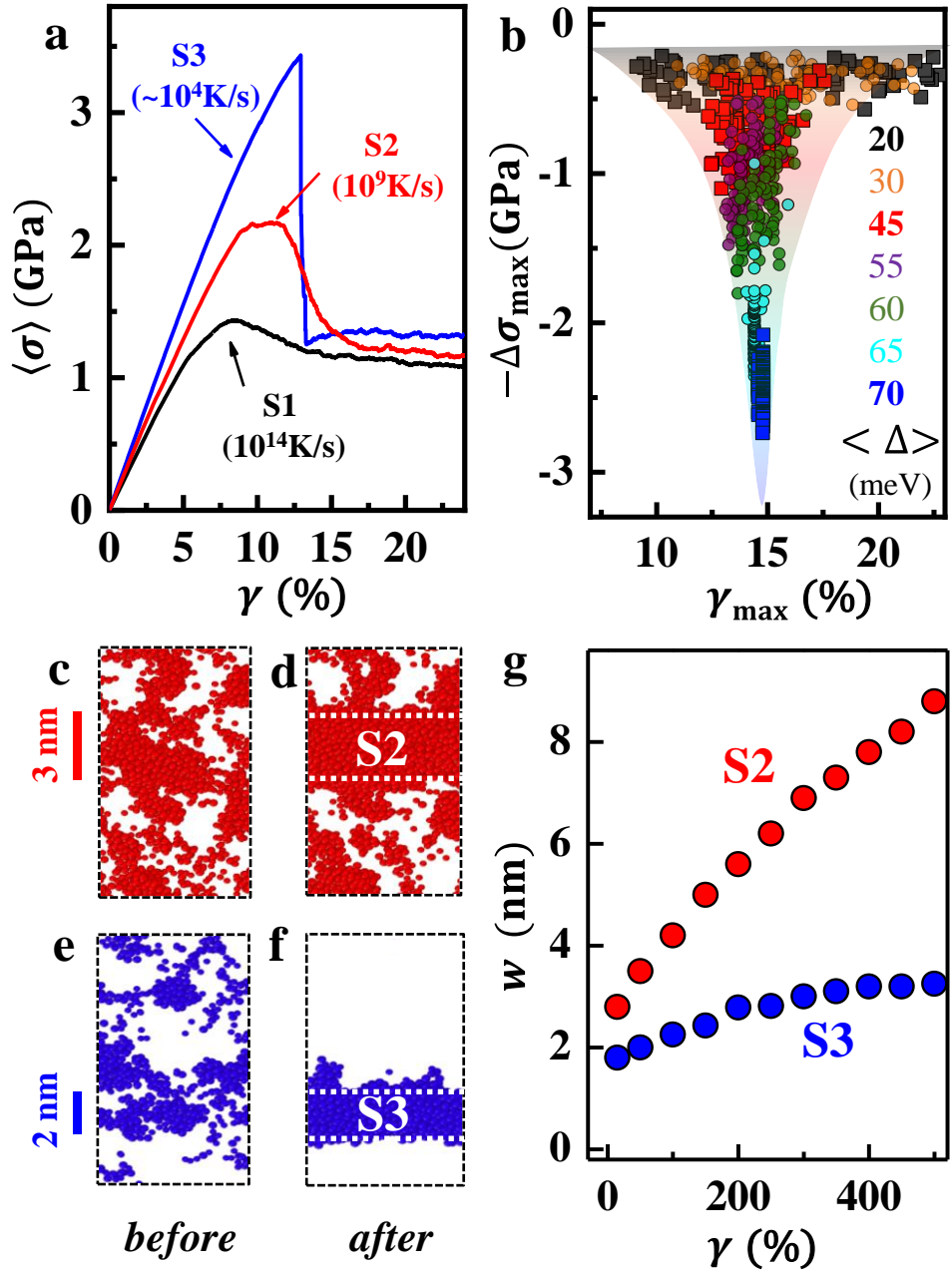

**Figure 2 | Annealing embrittlement revealed by the atomistic simulations. a,** Calculated stress-strain curves for samples **S1**, **S2**, and **S3** exhibit distinct yielding behaviors. **b,** The maximum stress drops $\Delta\sigma_{max}$ and the corresponding strains $\gamma_{max}$ at yielding of 100 replicas for each given $\langle\Delta\rangle$. **c-f,** Spatial distributions of local atomic rearrangements (atoms with $D^2_{min}>10$) just before (left panel) and after (right panel) yielding for samples **S2** and **S3**. **g,** The width ($w$) of shear band as a function of the external shear strain.

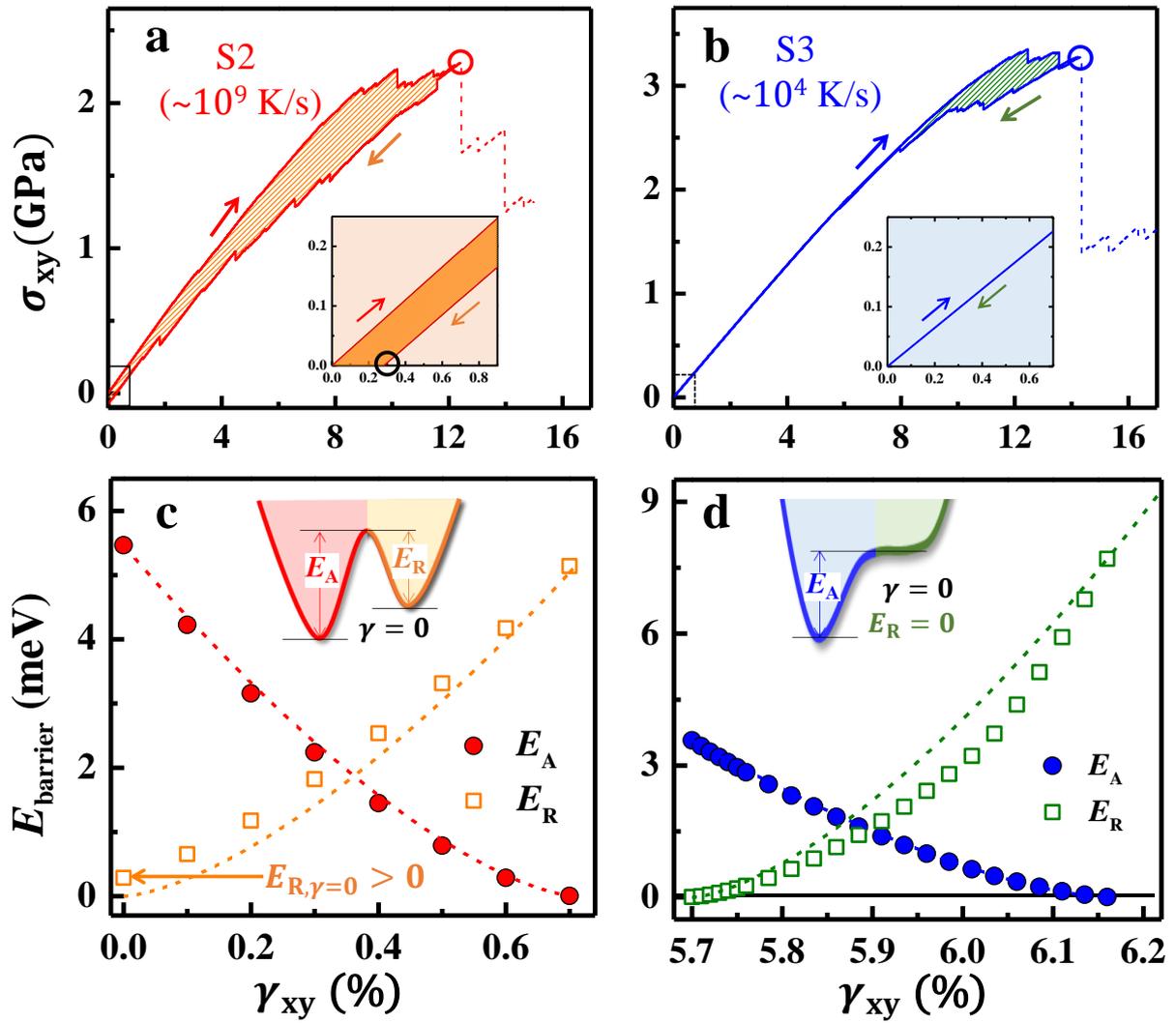

**Figure 3 | Thermally-activable versus strain-created ST events**. The calculated hysteretic cyclic stress-strain curves subjected to the external strain $\gamma_{xy}$ until yielding occurs in sample **S2** (**a**) and **S3** (**b**). The inset shows the enlarged view of the region close to $\gamma_{xy} = 0$. The calculated activation ($E_A$) and relaxation $E_R$ barriers against external strain $\gamma_{xy}$ for a typical event derived from the mechanical hysteresis in sample **S2 (c)** and **S3 (d)**. Note that the dashed curves are the prediction of the catastrophe theory: $E(\gamma) = A(\gamma_c - \gamma)^{3/2}$, where A is a constant pre-factor and $\gamma_c$ is the critical strain triggering a catastrophe event.

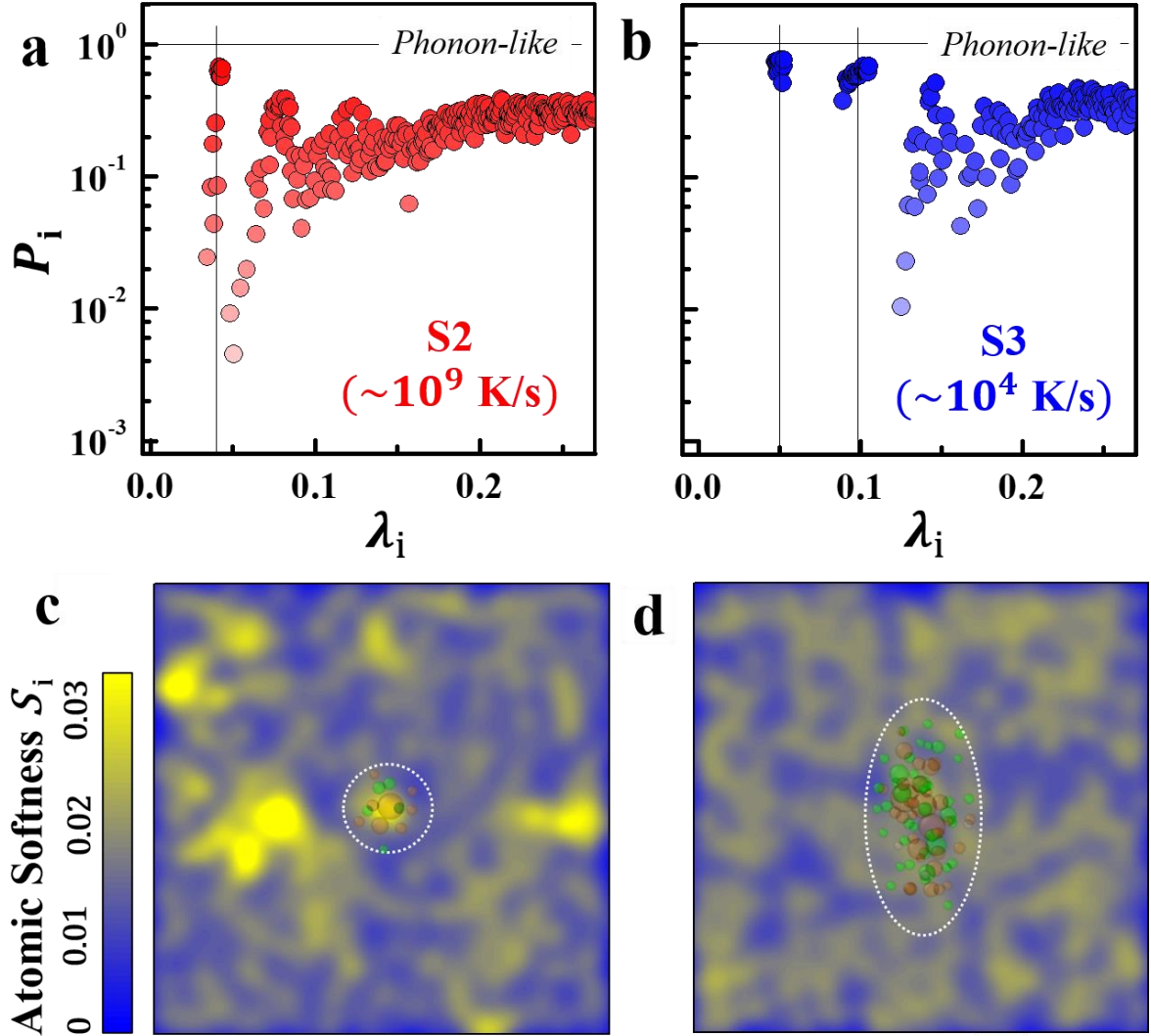

**Figure 4 | Characterization of thermal vibrations and soft spots**. **a-b**, Participation ratio $P_i$ of eigen modes in sample **S2** and **S3**, respectively. Spatial distributions of atomic softness $S_i = 1/N_m \sum_{i \leq N_m} |\vec{e_i}|$, which is calculated based on the vibrational modes of the inherent structure of sample **S2 (c)** and **S3 (d)**. Soft spots are colored in yellow, and atoms involved in the first event are encircled.

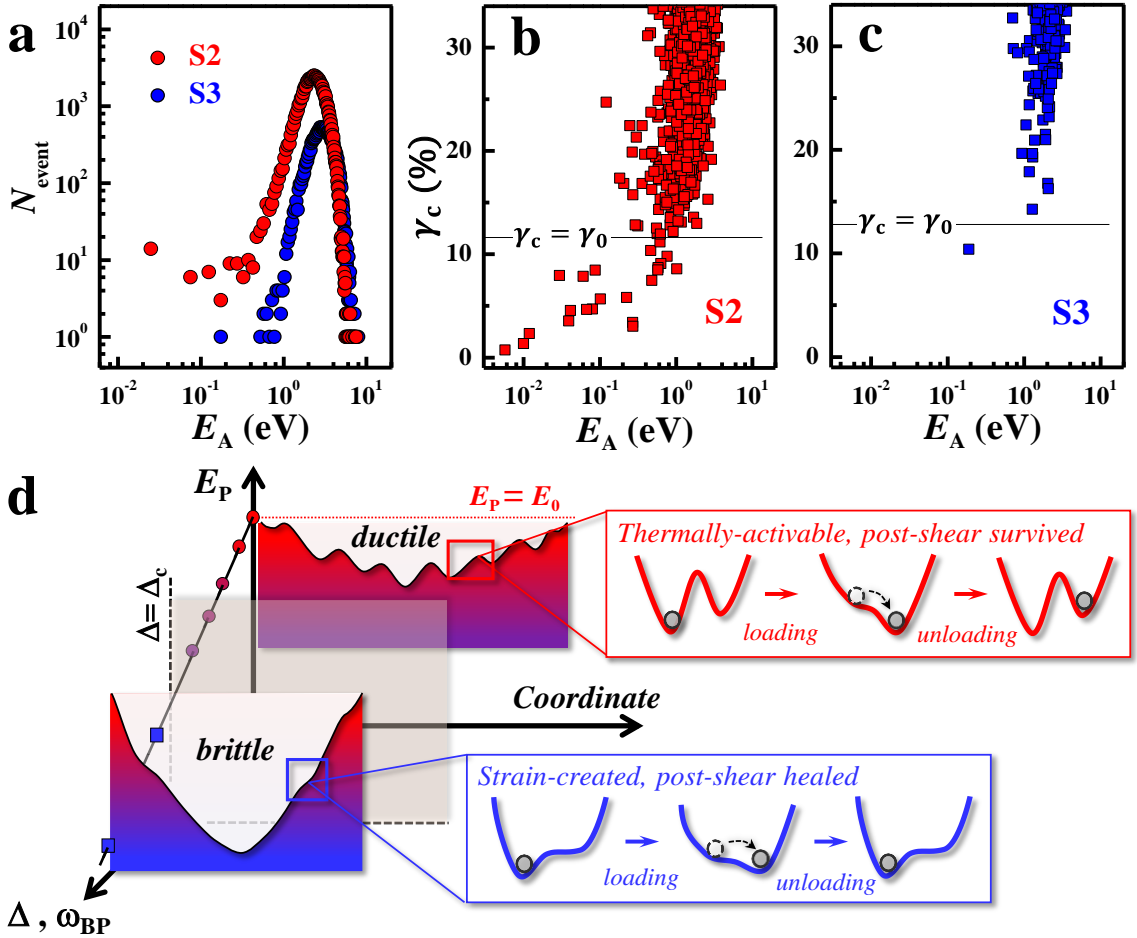

**Figure 5 | The spectra of activation barrier $E_A$ and the illustrated potential energy surface (PES). a,** Thermal activation barrier ($E_A$) spectra of sample **S**2 and **S**3. **b-c,** The predicted critical strains $\gamma_c$ of the insular thermally-activable events between neighboring sub-basins with various $E_A$ in samples S2 and S3, respectively. The yielding strains $\gamma_0$ of each sample are illustrated as thin solid lines. **d,** Schematic diagram of the local PES evolution that determines thermal-annealing embrittlement.

Supplementary Materials for

# Atomic Origin of Annealing Embrittlement in Metallic Glasses

Rui Su[1], Shan Zhang[2], Xuefeng Zhang[1,*], Yong Yang[3,4,*], Weihua Wang[5], Pengfei Guan[2,1,*]

[1]Institute of Advanced Magnetic Materials, College of Materials and Environmental Engineering, Hangzhou Dianzi University, Hangzhou 310018, P. R. China

[2]Beijing Computational Science Research Center, Beijing 100193, P. R. China

[3]Department of Mechanical Engineering, College of Engineering, City University of Hong Kong, Kowloon Tong, Kowloon, Hong Kong, China

[4]Department of Materials Science and Engineering, College of Engineering, City University of Hong Kong, Kowloon Tong, Kowloon, Hong Kong, China

[5]Institute of Physics, Chinese Academic of Science, Beijing 100193, P. R. China

**Corresponding authors.** XFZ (zhang@hdu.edu.cn), YY (yonyang@cityu.edu.hk) and PFG (pguan@csrc.ac.cn)

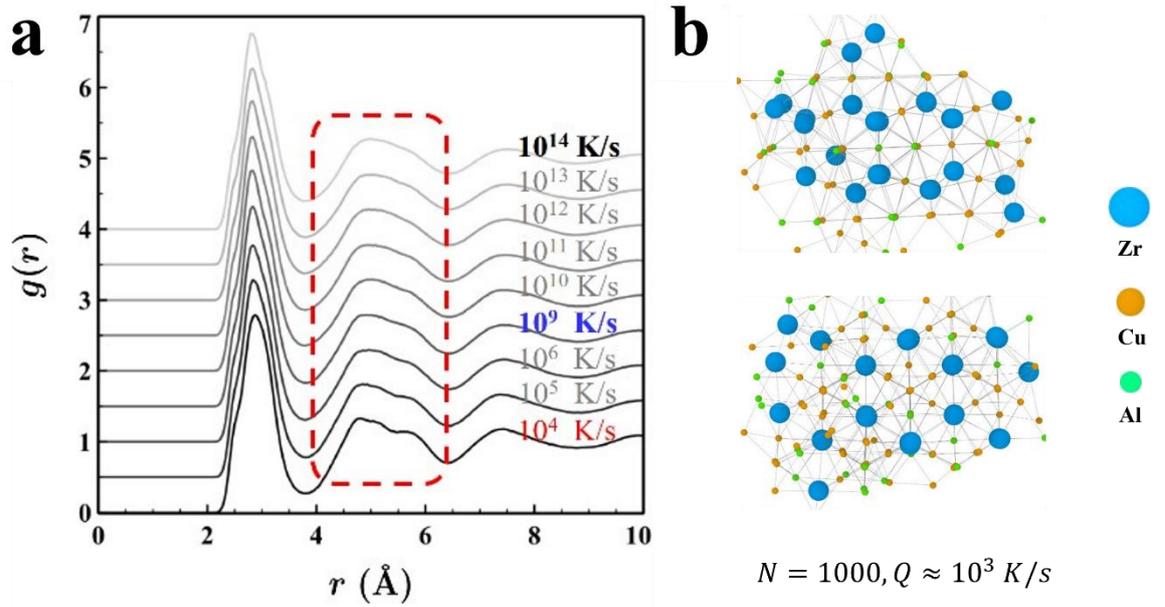

**Figure S1| Structure information of simulated metallic glass samples. a,** The calculated pair correlation functions $g(r)$ of various samples with different effective quenching rates, which show typical disorder behaviors. A noticeable difference is the splitting of second peaks at deeper annealing degree, which indicates enhanced structural medium-range order. **b**, Partial crystallization of samples with effective cooling rate $Q_e \sim 10^3\ K/s$.

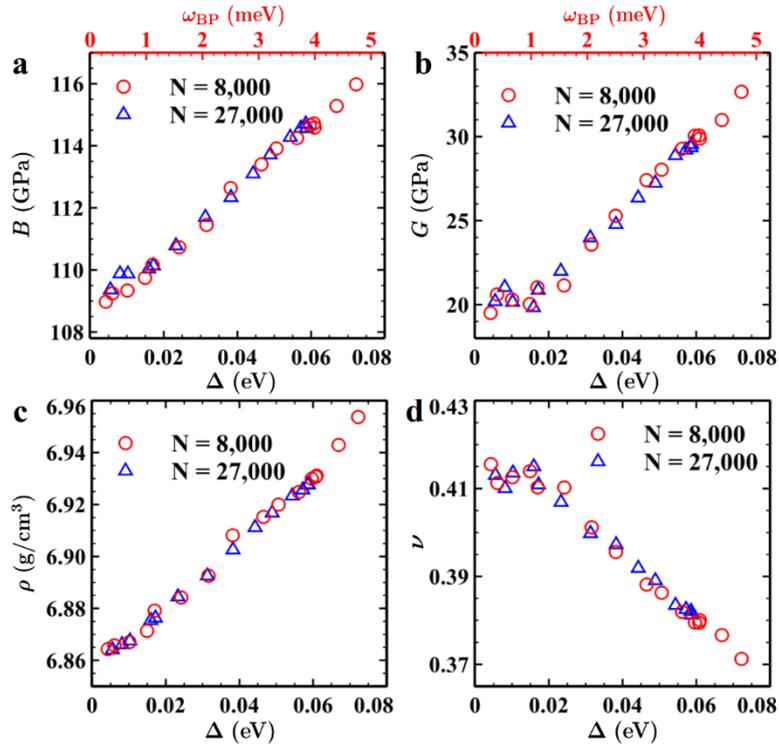

**Figure S2| Macroscopic properties of simulated MG samples with different annealing degree Δ for system sizes of N = 8,000 (red circles) and N = 27,000 (blue triangles). a, b:** Bulk and shear modulus with the respect to the annealing degree Δ; **c:** Mass density with the respect to the annealing degree Δ; **d:** Poisson's ratio with the respect to the annealing degree Δ. The corresponding Boson peak position $\omega_{BP}$ is shown as the upper axis in **a** and **b**, respectively. For all investigated macroscopic properties, no noticeable size effects can be observed.

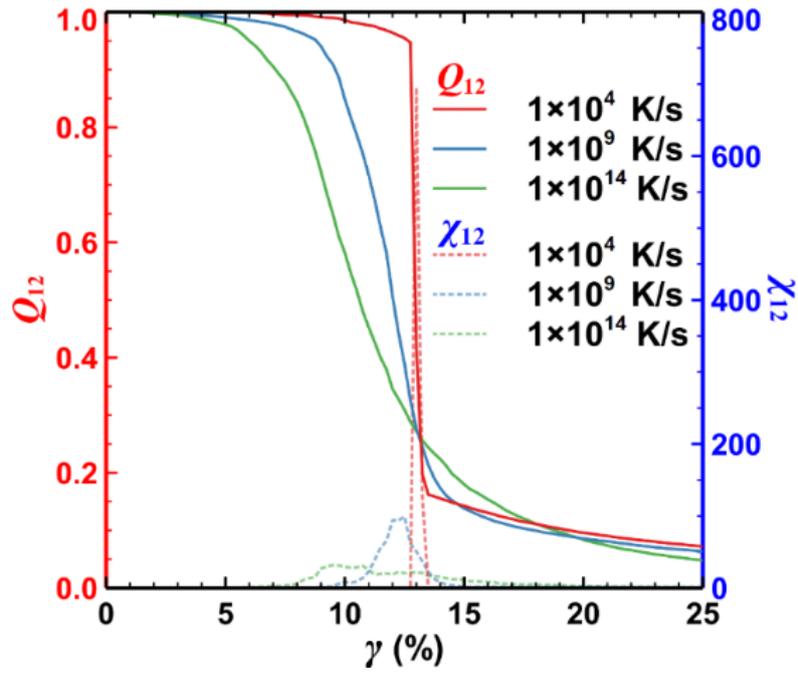

**Figure S3| The calculated overlapping function $Q_{12}$ and the dynamical susceptibility $\chi_{12}$ as a function of applied shear strain $\gamma$ for samples S1, S2 and S3.** A qualitative change from a smooth crossover (ductile-like behavior) for the poorly annealed S1 samples (Δ ≈ 20 meV/atom) and S2 samples (Δ ≈ 45 meV/atom) to a sharp discontinuity (brittle-like behavior) for the well-annealed S3 samples (Δ ≈ 70 meV/atom) can be observed. The sharp peak on the calculated $\chi_{12}$ of sample S3 confirms the brittle-like behavior.

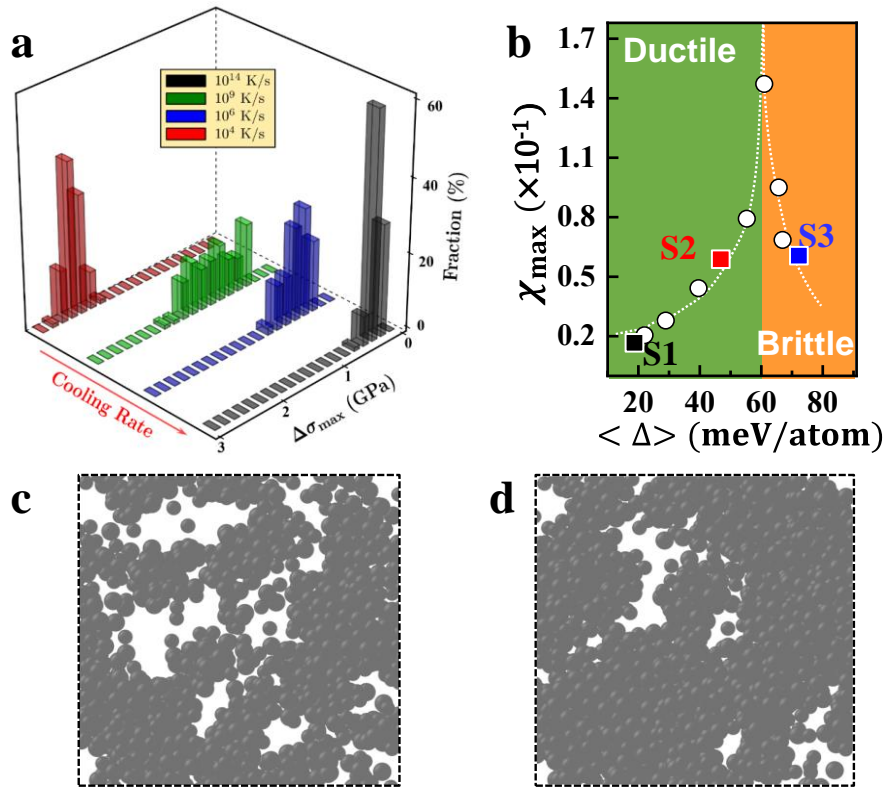

**Figure S4| Yielding behaviors of simulated MG samples. a:** The distribution of the maximum stress drops $\Delta\sigma_{max}$ of samples with a given $\Delta$. **b:** The calculated $\chi_{max}$ of samples with different $\Delta$. A sharp peak at a critical annealing degree: $\Delta_c \sim 60$ meV/atom ($Q_c \approx 10^6$ K/s) suggests a brittle-ductile transition. The snapshots of rearranged atoms (with $D^2_{min} > 5 Å^2$) in sample S3 before (**c**) and after (**d**) yielding.

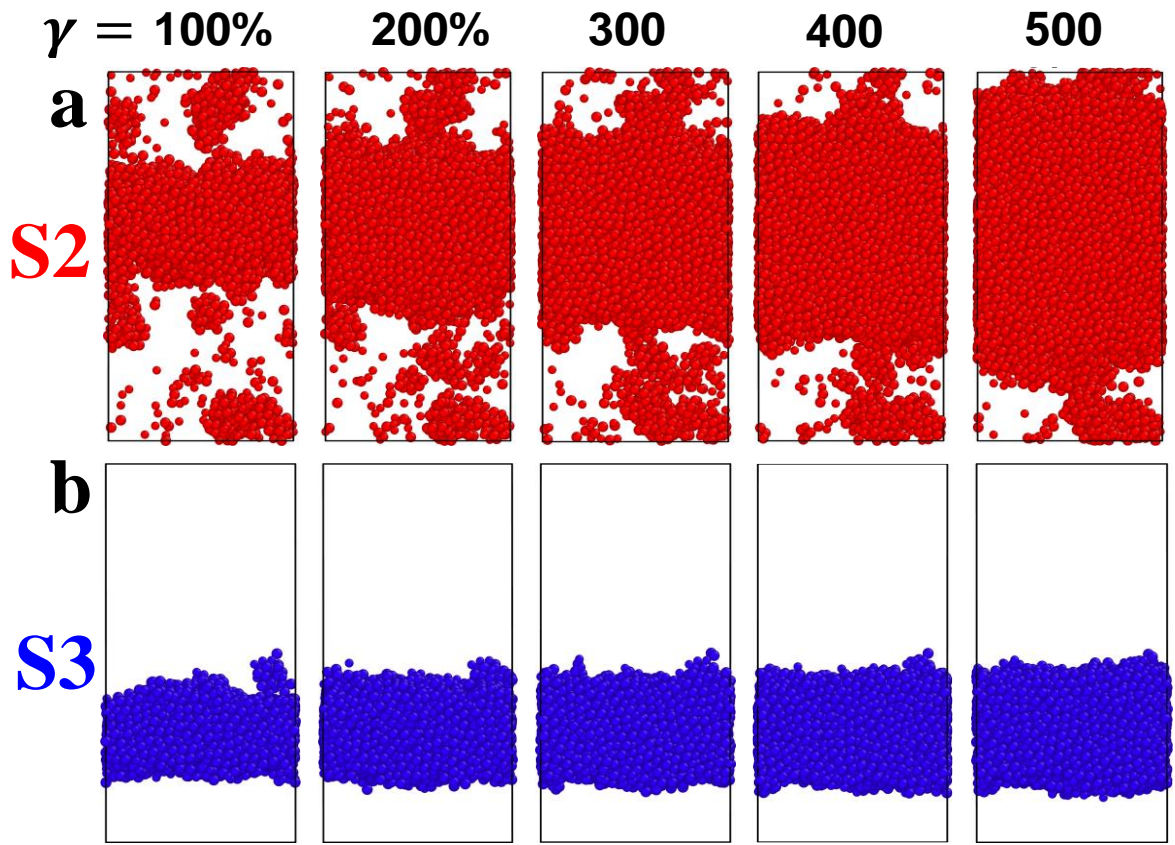

**Figure S5|** The broadening of shear band as the applied shear strain ($\gamma$) increases in sample S2 (**a**) and S3 (**b**).

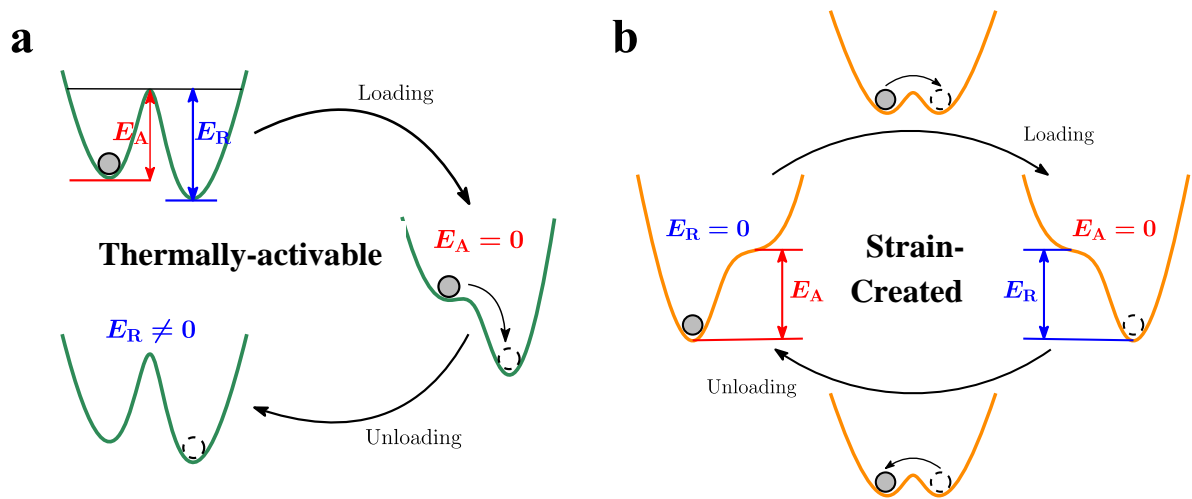

**Figure S6|** Illustrations of PES topography evolution according to the external shear strain for the thermally-activable (**a**) and strain-created (**b**) events, respectively.

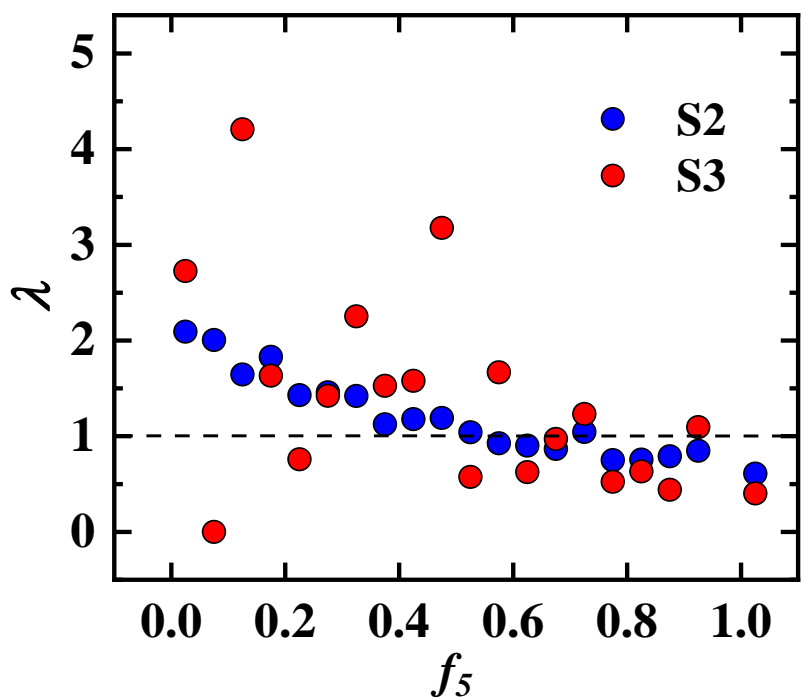

**Figure S7|** The calculated enhanced factor $\lambda$ for atoms with different local five-fold symmetrical parameter $f_5$. Here, $\lambda(f_5) > 1$ means that local structural rearrangements are more likely to occur at these atoms with the special local five-fold symmetry.

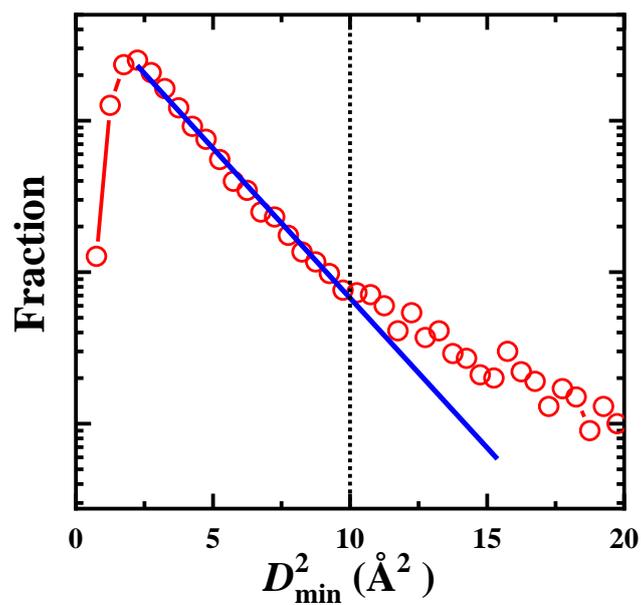

**Figure S8| The definition of the cutoff $D^2_{min}$ for selecting the rearranged atoms.** A cutoff $D^2_{min}$ of 5 Å² does not affect the main conclusion.